\documentclass[conference]{IEEEtran}

\pdfoutput=1

\usepackage{cite}
\usepackage{xcolor}
\usepackage{amsmath}
\usepackage{amssymb}
\usepackage{amsfonts}
\usepackage{graphicx}
\usepackage{algorithm}
\usepackage{algpseudocode}
\usepackage[hidelinks]{hyperref}
\usepackage{url}
\usepackage{dirtytalk}
\usepackage{enumitem}
\usepackage{pgfplots}
\usepackage{tikz}
\usepackage{tablefootnote}
\usepackage{threeparttable}
\pgfplotsset{compat=1.17}

\pgfplotsset{
  log x ticks with fixed point/.style={
      xticklabel={
        \pgfkeys{/pgf/fpu=true}
        \pgfmathparse{exp(\tick)}%
        \pgfmathprintnumber[fixed relative, precision=3]{\pgfmathresult}
        \pgfkeys{/pgf/fpu=false}
      }
  }
}

\begin{document}

\title{FiltPIM: In-Memory Filter for DNA Sequencing
% consider
%\\ BaCoFIM: Base-Count Filtering IN-Memory for DNA Sequencing
%\\ BacoFilt: Base-Count Filter for DNA Sequencing using Processing-In-Memory
%\\ IMBaCt: In-Memory Base-Count Filter for DNA Sequencing
%\\ BC-FoRM: Base-Count Filter for DNA Read Mapping
%\\ IMFoRM: In-Memory Filter for DNA Read Mapping
%\\ BaConMemory: Base-Count iN Memory Filter
%\\ In-Memory Base-Count Filtering for DNA Sequencing
%\\ PIMFilt: In-Memory Filtering for DNA Sequencing
%\\ In-Memory Pre-Alignmnet Filtering for DNA sequencing
}

\author{
\IEEEauthorblockN{Marcel Khalifa\IEEEauthorrefmark{1}, Rotem Ben-Hur, Ronny Ronen, Orian Leitersdorf, Leonid Yavits, and Shahar Kvatinsky} \IEEEauthorblockA{\emph{Viterbi Faculty of Electrical and Computer Engineering, Technion -- Israel Institute of Technology, Haifa, Israel}}
\IEEEauthorblockA{marcelkh@campus.technion.ac.il\IEEEauthorrefmark{1}, rotembenhur@campus.technion.ac.il, ronny.ronen@technion.ac.il, \\ orianl@campus.technion.ac.il, yavits@technion.ac.il, shahar@ee.technion.ac.il}%
}

\IEEEoverridecommandlockouts
\IEEEaftertitletext{\vspace{-15pt}}

\maketitle

\IEEEpubid{\begin{minipage}{\textwidth}\ \\[12pt] \centering \copyright 2022 IEEE. Personal use of this material is permitted.  Permission from IEEE must be obtained for all other uses, in any current or future media, including reprinting/republishing this material for advertising or promotional purposes, creating new collective works, for resale or redistribution to servers or lists, or reuse of any copyrighted component of this work in other works. \end{minipage}} 

% ---- Abstract ---- %
\begin{abstract}

Aligning the entire genome of an organism is a compute-intensive task. Pre-alignment filters substantially reduce computation complexity by filtering potential alignment locations. The base-count filter successfully removes over $\mbox{\boldmath $68\%$}$ of the potential locations through a histogram-based heuristic. This paper presents FiltPIM, an efficient design of the base-count filter that is based on memristive processing-in-memory. The in-memory design reduces CPU-to-memory data transfer and utilizes both intra-crossbar and inter-crossbar memristive stateful-logic parallelism. The reduction in data transfer and the efficient stateful-logic computation together improve filtering time by \boldmath $100$x compared to a CPU implementation of the filter. 

\end{abstract}

% ---- Keywords ---- %
% \begin{IEEEkeywords}
% \end{IEEEkeywords}

% ---- Introduction ---- %
\section{Introduction}
\label{sec:introduction}

\IEEEpubidadjcol

DNA sequencing, the process of reading the genome of a given organism, is the foundation of many scientific and medical discoveries. For example, human genome sequencing enables personalized medicine and early diagnosis of genetic diseases~\cite{AcceleratingGenomeAnalysis}.
A genome is a sequence of the bases [A, T, G, C], with the human genome being composed of approximately 3.2 billion bases. As reading the entire DNA sequence at once is infeasible for large-scale genomes, DNA sequencers typically extract sub-sequences called \textit{reads}. These reads are orders of magnitude shorter than the whole genome sequence, ranging from dozens of bases for \textit{short reads} to thousands of bases for \textit{long reads}~\cite{sequencing-Machines}. This work focuses on short reads.

DNA sequencing typically extracts numerous short reads, ranging from hundreds of millions to billions for the human genome. Then, through \emph{read mapping}, a step towards the genome assembly, the reads are aligned to a reference genome of a similar organism. Many previous works have accelerated read mapping on CPU~\cite{daily2016parasail, 10.1093/bioinformatics/btn025,langmead2009ultrafast}, GPU~\cite{ahmed2019gasal2, 9139808}, and FPGA \cite{10.1093/bioinformatics/btx342,6822570}. Standard hardware, however, suffers from the massive amount of data transfer between the memory and the processing unit (memory wall bottleneck). Therefore, \textit{processing-in-memory} (PIM) platforms have gained interest. Such platforms inherently support parallel logic on sets of data residing in memory without the need to read the data, thereby reducing the memory wall bottleneck~\cite{mMPU}.

\begin{figure}[t]
    \centering
    \includegraphics[width=\linewidth]{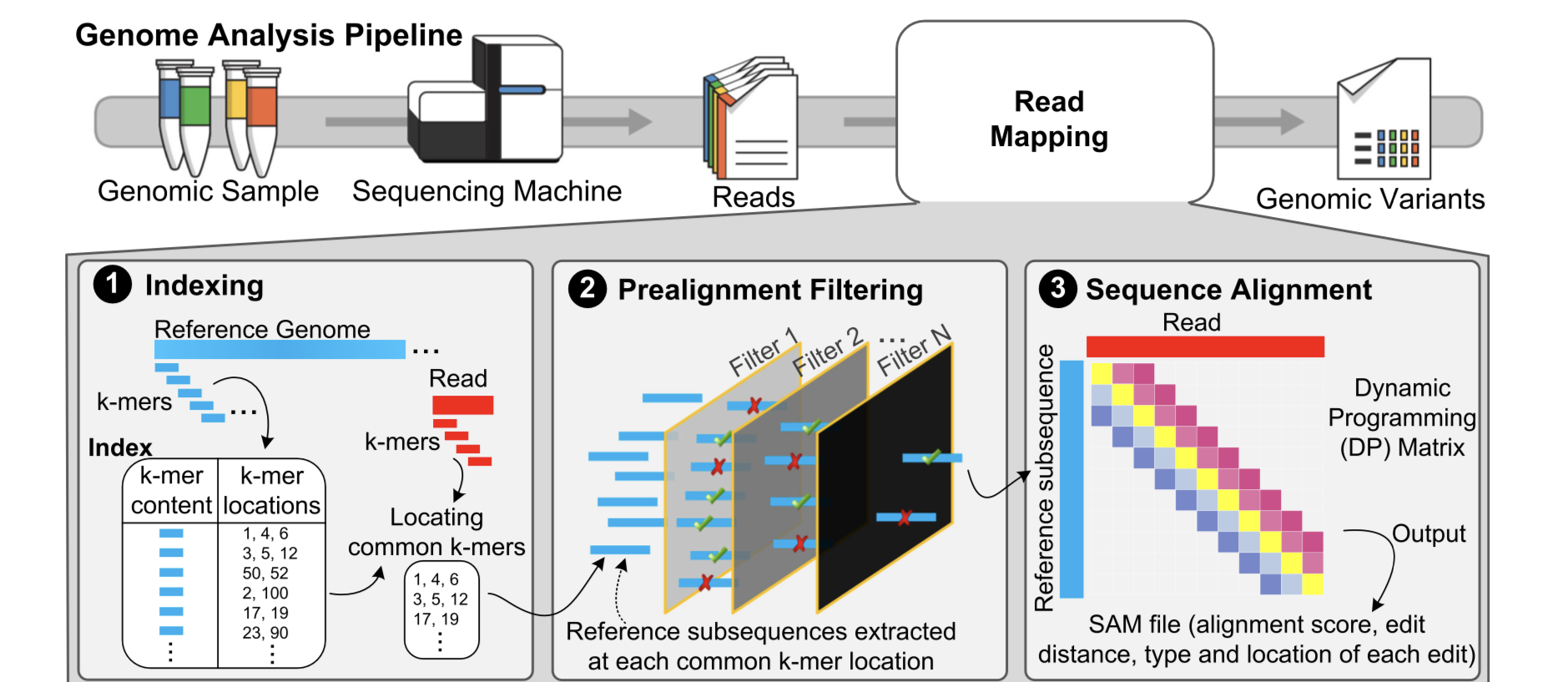}
    \caption{DNA sequencing pipeline, demonstrating the \emph{indexing}, \emph{prealignment filtering}, and \emph{sequence alignment} stages. Figure adapted from~\cite{AcceleratingGenomeAnalysis}.\\
    \copyright \ 2020, IEEE.}
    \label{fig:DNAconcept}
    \vspace{-10pt}
\end{figure}

Read mapping can be divided into the \emph{indexing}, \emph{pre-alignment filtering}, and \emph{sequence alignment} stages, as shown in Figure~\ref{fig:DNAconcept}. To guide the mapping process, a full reference genome of a similar organism is utilized. Initially, the indexing stage generates potential locations on the reference genome for each read. We focus on the pre-alignment stage, where the similarity between a DNA read and reference fragments is evaluated via heuristics. This stage filters false-positives for the computationally-intensive optimal sequence alignment.

Pre-alignment filtering aims to remove potential locations with edit distance above a certain predefined edit threshold (\textit{eth}) compared to the read. Edits may be substitutions, insertions, or deletions. They are allowed due to genome variations among different organisms as well as sequencing errors.
The base count filter~\cite{AcceleratingGenomeAnalysis} is a pre-alignment filtering algorithm that discards more than 68\% of the potential locations by using a histogram-based heuristic for similarly comparison. Crucially, the base-count filter does not harm the sensitivity of the mapper as it does not discard true locations. 

\IEEEpubidadjcol

The base count filter operates on massive amounts of data, which makes it suitable for PIM. In this paper, we present FiltPIM, a memristive processing-in-memory accelerator that efficiently implements this filter. The PIM capabilities of FiltPIM are based on memristive stateful logic~\cite{MAGIC}. Stateful logic provides inherent inter-crossbar and intra-crossbar parallelism. By reducing data transfer and exploiting both types of parallelism, FlitPIM reduces filtering time by 100x compared to the CPU implementation of the filter.

% ---- Background ---- %
\section{Memristor-Aided Logic (MAGIC)}
\label{sec:MAGIC}

\begin{figure}
    \centering
    \includegraphics[width=\linewidth]{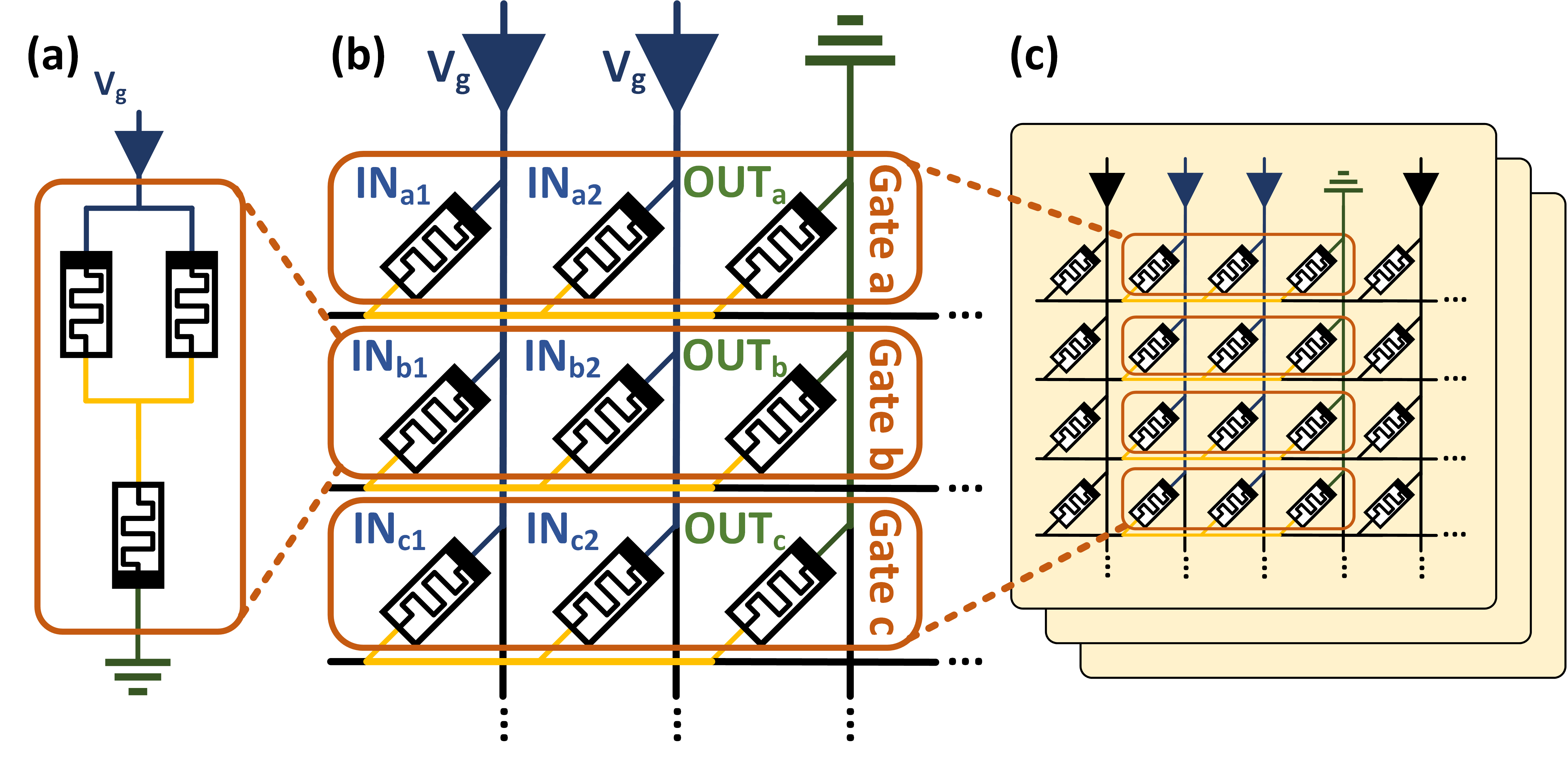}
    \caption{{(a) MAGIC NOR gate. (b) Parallel mapping of the MAGIC NOR gate to crossbar array rows, and (c) parallel computation across crossbars.}}
    \label{fig:XB}
    \vspace{-15pt}
\end{figure}

Memristive stateful logic is a PIM concept for representing data with resistance and performing logic operations on the resistances using the same cells in a memristive memory crossbar array~\cite{PATMOS}. One of the classic stateful logic techniques is MAGIC~\cite{MAGIC}. MAGIC implements logic gates using memristor cells within the same row (or within the same column) of a memristive crossbar array, see Figure~\ref{fig:XB}. The inputs of the MAGIC gate are the states (resistance) of the input memristors prior to the computation, and the output is the resistance of the output memristor after computation. MAGIC NOR is executed in two steps (clock cycles): (1) initializing the output memristor to logical '1' (low resistance), (2) applying a voltage $V_g$ across the gate. 

MAGIC supports inherent parallelism as the same in-row gate can be performed in parallel across multiple rows (see Figure~\ref{fig:XB}(b)) and across multiple crossbars (see Figure~\ref{fig:XB}(c)).
Table~\ref{tab:MAGICOperations} lists different operations that are implemented using MAGIC NOR gates and specifies the number of cycles they require per single bit operands (initialization cycles are not included, as different initialization can be executed in parallel in the same clock cycle~\cite{SIMPLER}). 
All operations, except \textit{Popcount} (count 1's in a given column), are straightforward. 

\begin{figure}
    \centering
    \includegraphics[width=3.1in,trim={1cm 2.5cm 0.5cm 1cm},clip=true]{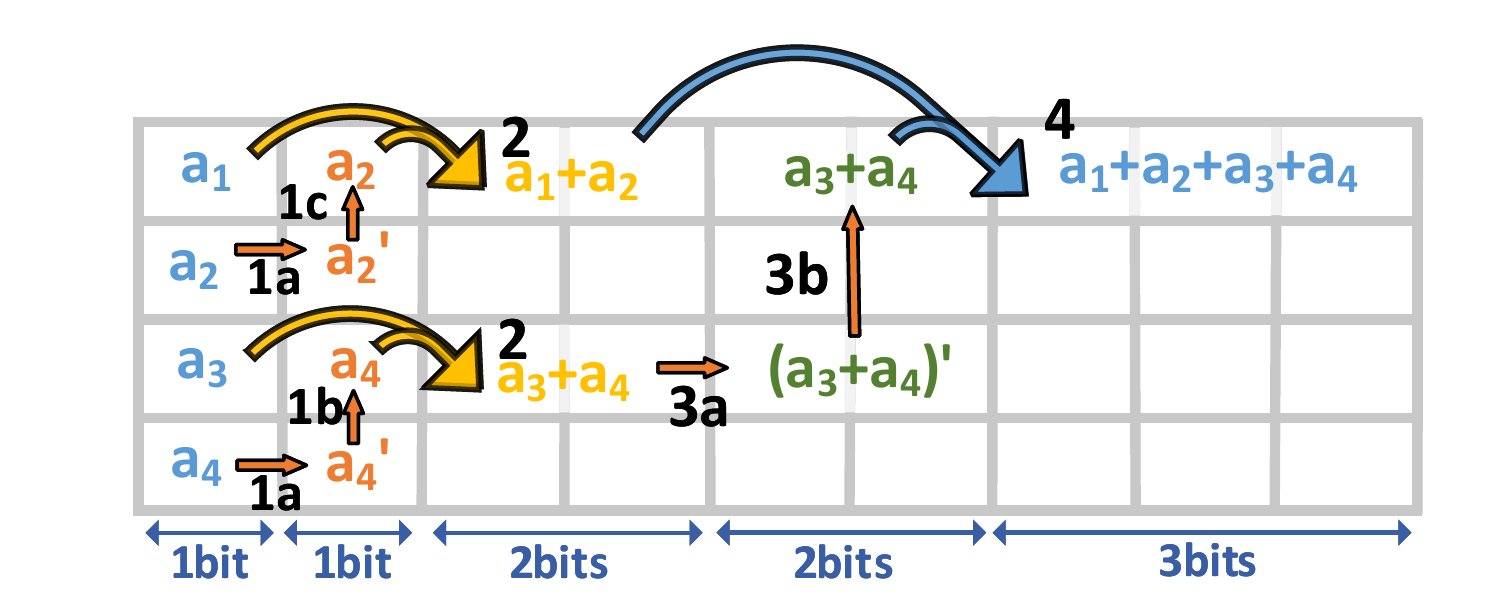}
    \caption{Calculating the popcount of a 4-bit column using MAGIC NOR/NOT.}
    \label{fig:popcount}
    \vspace{-10pt}
\end{figure}

Popcount is a special case of the reduction algorithm introduced by Ronen~\textit{et al.}~\cite{bitlet}. The algorithm is based on a recursive-tree technique that pairs up numbers and accumulates them in parallel, as seen in Figure~\ref{fig:popcount}. 
The first iteration is to add each two vertically adjacent numbers by aligning them in the same row (steps 1a, 1b, 1c in Figure~\ref{fig:popcount}), and then summing them (step 2). We iteratively continue with summing the vertically adjacent results (steps 3a, 3b, 4). In the last iteration, we will the sum of all bits residing at the end of the first row. A popcount of 100 bits requires $414$ cycles.

\begin{table}[t]
    \centering
    \caption{MAGIC Operations}
    \begin{tabular}{|c|c|c|}
        \hline
        Operation & Cycles/bit & Notes \\
        \hline
        NOR/NOT & $1$ & NOT is a 1-input MAGIC NOR \\
        \hline
        Copy & $2$ & NOT(NOT(cell)) \\
        \hline
        Half Adder & 5 & \cite{HA} \\
        \hline
        $N$-bit Adder & $9\cdot N +1$ & \cite{bitlet} \\ 
        \hline
        $N$-bit Subtractor & $9\cdot N +1$ & \cite{FS} \\ 
        \hline
        Popcount for 100 bits & $414$ & Section~\ref{sec:MAGIC} \\
        \hline 
        $N$-bit MUX($X,Y,sel$) & $4\cdot N$ & $\left((x_i+sel')' + (y_i+sel)'\right)'$\\
        \hline
        
    \end{tabular}
    \label{tab:MAGICOperations}
\end{table}

\section{Base Count Filtering}
\label{sec:BaseCountFiltering}

%Pre-alignment filtering aims to quickly determine whether a potential location is worth further analysis. It substantially reduces the number of locations to be examined using computationally-heavy optimal alignment algorithms, such as Smith-Waterman, in the Sequence Alignment stage, thereby reducing the overall alignment time.

The base-count filter introduced by Wendi \textit{et al.}~\cite{filter}, aims to discard locations that have more than $eth$ (edit-threshold) edits. The filter is based on comparing the base-count histograms of the read and the potential location. For each base type $B \in [A, T, G, C]$, the base error is defined as the absolute value of the count differences for the base. These base errors are accumulated to receive a single error. If that error is greater than $2\cdot eth$, then the read and the potential location must have an edit distance of at least $eth$ and thus the potential location is discarded. As histograms are not influenced by permutation, then the heuristic does not remove all locations with more than $eth$ errors.

CPU implementations of the filter, such as GASSST~\cite{GASSST}, limit the comparison to a sub-sequence of the read to reduce computation complexity.
In FiltPIM, we exploit the parallelism of the memristive crossbars to compare base count for the entire read, thus ensuring higher precision (higher number of discarded locations) while reducing the execution time.

To evaluate the efficiency of the filter, we incorporate a CPU implementation of the algorithm into \textit{mrFAST}~\cite{mrFAST}, a state-of-the-art read mapping tool. Various human-genome data-sets with 100-base reads were considered: \textit{ERR240726\_1}, \textit{ERR240727\_1}, and \textit{ERR240730\_1}\footnote{Available at www.ebi.ac.uk/ena/browser.}. The filter successfully discards more than 68\% of \textit{all potential locations}, while not affecting the sensitivity of the mapper as no true locations were discarded.

\algnewcommand\algorithmicforeach{\textbf{for each}}
\algdef{S}[FOR]{ForEach}[1]{\algorithmicforeach\ #1\ \algorithmicdo}

\begin{figure*}
    \centering
    \includegraphics[width=6.5in, trim={0.1cm 5.75cm 0.1cm 2.5cm},clip=true]{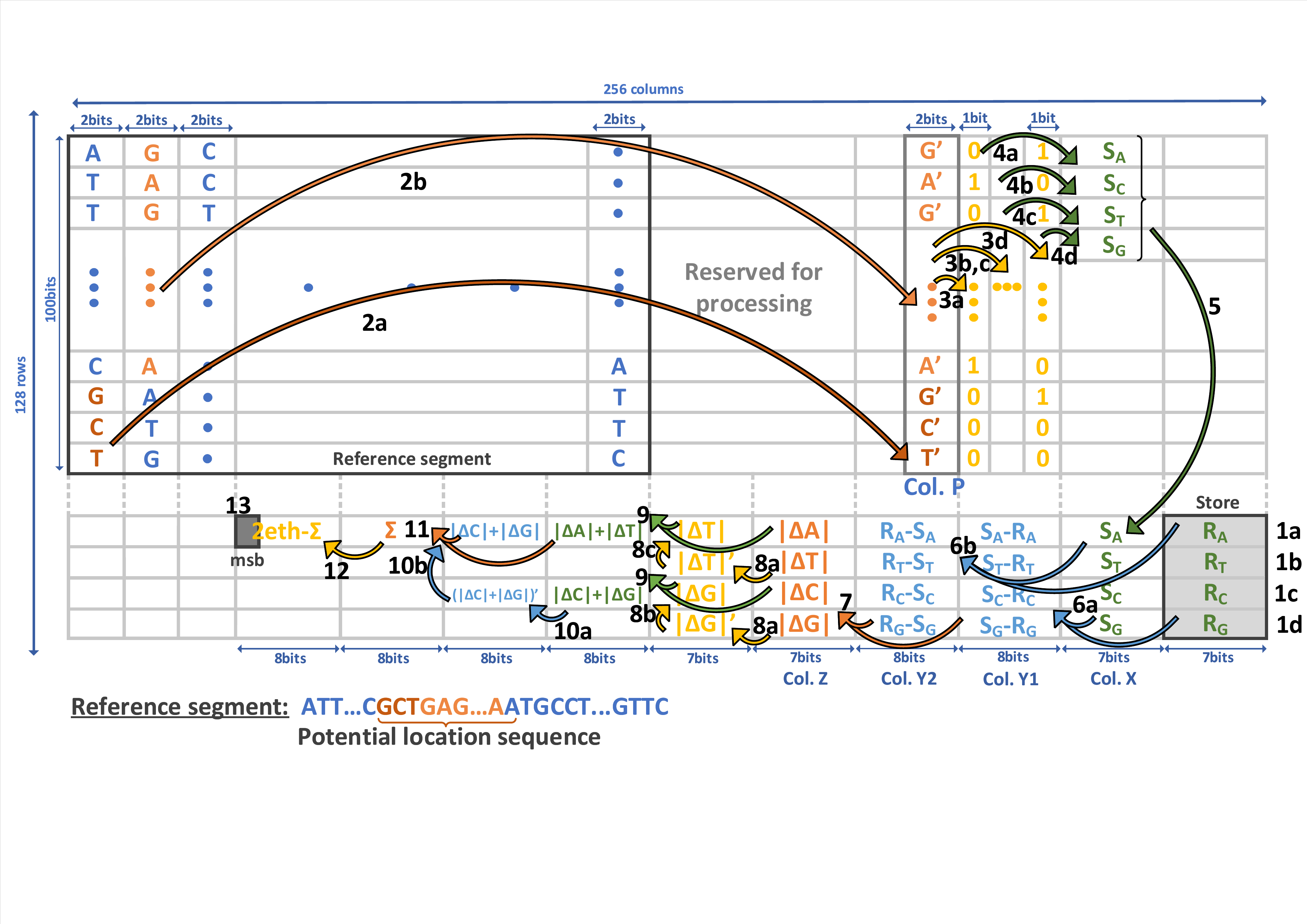}
    \caption{Base-count filtering within a memristive memory array. The steps in black are explained in detail in Section~\ref{sec:Base-Count Filtering Within a Memristive Memory MAT}. }
    \label{fig:inMAT}
\end{figure*}

% ---- Base-Count Filtering Within mMPU ---- %
\section{Base-Count Filtering Within a Memristive Memory Array}
\label{sec:Base-Count Filtering Within a Memristive Memory MAT}

In this section, we demonstrate an implementation of the filtering algorithm within a single memristive crossbar array. The array is structured to support reads of length up to 100-bases, yet it can be expanded to support other lengths.

We consider 128x256 arrays. 128 rows allow pre-storage of 100 bases of a reference genome within a column while reserving the remaining cells for temporary storage. The 256 columns enable pre-storage of a significant portion of the reference genome within each array: $65\cdot 100 = 6500~bases$, implying 130 columns as two adjacent bits are used to represent each base. An array is selected to participate in the filtering only when a fragment of the sequence containing the potential location fully resides within that array. The selected array receives the base-count histogram of the read (the number of occurrences of each base type [A, G, C, T]) and the exact potential location to be checked (a total of $4\times 7+32=60\ bits$), and returns \emph{true} if the location is to be discarded (returns $32\ bits$ location + $1\ bit$ result). To check the location, a series of operations inside the array are executed (see Figure~\ref{fig:inMAT}): 

\begin{itemize}
    \item \emph{Step 1:} For each base type \textit{B}, the number of occurrences for \textit{B} in the read, $R_B$, is stored to reserved cells in the array. 2 cycles to store each value. %(1~cycle to write '0' and another to write '1')
    \textbf{\textit{8~cycles in total}}.
    
    \item \emph{Step 2:} A NOT operation is performed on the reference genome fragment of length $read\_length$, starting at $potential\_location$, to copy it (inverted) to column \textit{P}. It is done in two steps, as the bases of the sub-sequence may span over two different 2-bit columns. Note that the bases will not overlap horizontally (as the number of bases in each column is each to the read length). \textbf{\textit{4~cycles in total}}.
    
    \item \emph{Step 3:} For each base type \textit{B}, \textit{P} is compared against $NOT(B)$.
    Each base is represented by two bits, as follows: A="00", T="01", G="10", C="11". 
    For each type \textit{B}, a different function is performed to compare a base against it, as listed in Table~\ref{tab:compareBases}.
    \textbf{\textit{6~cycles in total}}.
    
    \item \emph{Step 4:} For each base type \textit{B}, the matches of \textit{B} in \textit{P} are counted by popcounting the 100-bit columns computed in step 3. \textbf{\textit{1656~cycles in total}}.
    
    \item \emph{Step 5:} For all base types, the base counts in a potential location's sequence and in the read are aligned by copying the results from step 4 to column \textit{X}. Two NOTs are performed to copy each result. \textbf{\textit{8~cycles in total}}.
    
    \item \emph{Step 6:} For all base types, in parallel, perform  $S_B - R_B$ and then compute their 2's complement ($R_B - S_B$) by inverting all bits followed by performing Half Adder\textit{ (HA)} \textit{8 times} along the bits. \textbf{\textit{111~cycles in total}}.
    
    \item \emph{Step 7:} For all base types, in parallel, calculate $|S_B - R_B|$ by choosing the non-negative value between both values calculated in step 6.
    To do this, we apply a MUX operation (see Table~\ref{tab:MAGICOperations}), where the \textit{most significant bit} (msb) of the result in column $Y_1$ is the selector. The result in column $Y_2$ is \textit{X} and the result in column $Y_1$ is \textit{Y}. If, for example, the selector is '0' ($msb_{(S_B - R_B)}='0'$), we conclude that $(S_B - R_B) \geqslant	0 $ and that this value shall be chosen. \textbf{\textit{28~cycles in total}}.
    
    \item \emph{Steps 8-11:} Addition is performed between all differences in counts (four values in column Z). This is performed in two iterations. In each iteration each two vertically adjacent values are aligned in the same row, and then full-adder \textit{(FA)} is performed. \textbf{\textit{153~cycles in total.}}
    
    \item \emph{Step 12:} The sum of all differences in count is subtracted from the pre-defined constant ($2\cdot eth$).
    \textbf{\textit{72~cycles in total}}.
    
    \item \emph{Step 13:} Read out the msb of the final result of step 12.  \textbf{\textit{1~cycle in total.}}
    If the msb of the result is '1', meaning the sum of differences in counts is greater than twice the error threshold, then the potential location is discarded.
\end{itemize}

In summary, using in-crossbar array operations, the validity of a potential location for a given read can be checked in \textbf{\textit{less than 3000 MAGIC-NOR cycles}} including initialization cycles (less than 2000 cycles without initialization cycles).

\begin{table}[t]
    \centering
    \caption{Base comparisons}
    \begin{tabular}{|c|c|c|c|}
        \hline
        Compare \say{ab} & \#Cycles & Function & Notes \\
         against & & & \\
        \hline
        NOT(A) & 3 & NOR(NOR(a), NOR(b)) & \\
        \hline
        NOT(T) & 1 & NOR(NOR(a), b) & NOR(a) already \\
         & & & computed \\
        \hline
        NOT(G) & 1 & NOR(a, NOR(b)) & NOR(b) already \\
         & & & computed \\
        \hline
        NOT(C) & 1 & NOR(a, b) & \\ 
        \hline
        
    \end{tabular}
    \label{tab:compareBases}
\end{table}

\section{Evaluation}
\label{sec:Evaluation}

To evaluate FiltPIM, we compare its performance against a CPU-based implementation of an equivalent base-count filter. The advantage of FiltPIM originates from its massive intra-crossbar and inter-crossbar parallelism, and from the reduced CPU-to-memory data transfer.
%the main advantage of the in-memory filtering stems from using many crossbar arrays, where each crossbar can check one potential location in parallel to all others. There is also a gain from the reduced data transfer, but it is less significant. 

The human genome consists of approximately 3.2 billion bases. We pre-store the genome in 500,000 crossbars, each containing 64+1=65 read-size (100 bases) \textit{fragments}. The genome portion within a crossbar has dimensions $100 \ rows \times 130 \ columns$. The first and the last fragments in each crossbar overlap with the neighboring crossbars to guarantee that each potential-location fragment resides in a single crossbar.

A set of human-genome reads (\textit{ERR240727\_1}) is considered. When aligning the reads against a human genome reference (\textit{humanG1Kv37}), the indexing stage of mrFast provides a total of 46 billion potential locations.
%old
%Assuming potential locations are distributed uniformly, each crossbar is accessed on average $46G/500K = 92,000$ times. %Accessing each crossbar 100,000 times results in checking over 99\% of the locations, passing less than $1\%$ to the sequence aligner without checking. 

%new
%In each crossbar, computing up to 5 times the average locations per crossbar ($5\cdot46G/500K = 460K$) Will cover almost all locations.\footnote{This decreases the filter's efficiency by less than 0.5\% (We examined the distribution of all locations against the arrays).}

To evaluate the improvement of FiltPIM over CPU, we developed an optimized standalone implementation for the base-count filter, and measured the latency for checking 46 billion potential locations. The tool was executed on an Intel(R) Xeon(R) CPU E5-2683 v4 at 2.1 GHz, with 256GB of DRAM, 2400MHz DDR4, 2x 1TB HD and 480GB SSD.
We observed total latency of approximately $7360$ seconds, of which $70\%$ are for data transfer and $30\%$ for computation.

We analyzed FiltPIM performance using the algorithm from Section~\ref{sec:Base-Count Filtering Within a Memristive Memory MAT}. We assumed the same CPU-to-memory interface (2400MHz DDR4). The actual data transfer rate may depend on the full architecture of FiltPIM, which is out of scope for this paper. Therefore, we assumed 10 GB/s, about half of the peak performance rate for DDR4 2400 (19.2 GB/s). Table~\ref{tab:results} summarizes the results.

\begin{table}[t]
    \centering
\begin{threeparttable}
    \caption{Evaluation Results for FiltPIM}
    \label{tab:results}
    \begin{tabular}{|c|c|c|c|}
        \hline
        & \textbf{\emph{Compute}} & Notes & FiltPIM \\
        \hline
        \hline
        a & Cycle time (ns) & \cite{bitlet} & $10$  \\
        \hline
        b & PIM Crossbars &  & $500,000$ \\
        \hline
        c & Latency per iteration (cycles) & & $3,000$ \\
        \hline
        d & Latency per iteration (ns) & $a \times c$ & $30,000$\\
        \hline
        e & \# Potential locations & & $46\cdot 10^9$ \\ 
        \hline
        f & \# Iterations & $5 \times (e/b)$ \tnote{1} &
        $460,000$   \\
        \hline
        g & Total latency (sec) & $d\times f$ & $13.8$\\
        \hline
        \hline
        & \textbf{\emph{Transferred Data}} &  &  \\
        \hline
        \hline
        h & Transfer per potential location (B) & data in\&out & $8+5=13$ \\
        \hline
        i & Data transfer rate (GB/sec) & & $10$  \\
        \hline
        j & Total data transferred (GB) & $h\times e$ & $598$ \\
        \hline
        k & Data transfer latency (sec) & $j/i$ & $59.8$ \\
        \hline
        \hline
        &\textbf{ \textit{Total Time (sec)}} & $k+g$ & \emph{$73.6$} \\
        \hline
    \end{tabular}
   \begin{tablenotes}
   \item[1] Iterating 5 times the average locations per crossbar ($5\cdot46G/500K = 460K$) in each crossbar, covers over 99\% of all locations, decreasing  the filter's efficiency by less than 0.7\% (inferred from the locations distribution among the crossbars).
   \end{tablenotes}
\end{threeparttable}
\vspace{-12pt}
\end{table}

The results show that FiltPIM is faster than the CPU on both the computation time (160x) and the data transfer time (86x). In total, it reaches a speedup of 100x over the CPU.

\begin{figure}[!t]
\centering 
\begin{tikzpicture}
\begin{loglogaxis}[
    xlabel={\# of Crossbar Arrays},
    ylabel={Latency [Seconds]},
    xtick={0,10, 200, 10000, 500000},
    ytick={1e1, 1e2, 1e3, 1e4, 1e5, 1e6, 1e7},
    ymin=10^1.5,
    ymax=10^7,
    legend pos=north east,
    every axis plot/.append style={ultra thick},
    width=3.5in,  %3.2
    height=2.6in,  %2.1
    log x ticks with fixed point
]
    
\addplot[
    orange,dashed, ultra thick,
    mark=none,
    const plot,
    empty line=jump,
    ]
    coordinates {(1, 7360)(500000, 7360)
    };

\addplot[
    color=blue,
    ]
    coordinates {(1,1380059.000000)(1001,1437.621379)(2001,748.655172)(3001,518.846718)(4001,403.913772)(5001,334.944811)(6001,288.961673)(7001,256.114698)(8001,231.478440)(9001,212.316298)(10001,196.986201)(11001,184.443142)(12001,173.990417)(13001,165.145681)(14001,157.564388)(15001,150.993867)(16001,145.244610)(17001,140.171696)(18001,135.662408)(19001,131.627756)(20001,127.996550)(21001,124.711157)(22001,121.724422)(23001,118.997391)(24001,116.497604)(25001,114.197792)(26001,112.074882)(27001,110.109218)(28001,108.283954)(29001,106.584566)(30001,104.998467)(31001,103.514693)(32001,102.123652)(33001,100.816915)(34001,99.587042)(35001,98.427445)(36001,97.332269)(37001,96.296289)(38001,95.314834)(39001,94.383708)(40001,93.499138)(41001,92.657716)(42001,91.856361)(43001,91.092277)(44001,90.362924)(45001,89.665985)(46001,88.999348)(47001,88.361077)(48001,87.749401)(49001,87.162691)(50001,86.599448)(51001,86.058293)(52001,85.537951)(53001,85.037245)(54001,84.555082)(55001,84.090453)(56001,83.642417)(57001,83.210102)(58001,82.792693)(59001,82.389434)(60001,81.999617)(61001,81.622580)(62001,81.257706)(63001,80.904414)(64001,80.562163)(65001,80.230443)(66001,79.908774)(67001,79.596708)(68001,79.293819)(69001,78.999710)(70001,78.714004)(71001,78.436346)(72001,78.166400)(73001,77.903851)(74001,77.648397)(75001,77.399755)(76001,77.157656)(77001,76.921845)(78001,76.692081)(79001,76.468133)(80001,76.249784)(81001,76.036827)(82001,75.829063)(83001,75.626306)(84001,75.428376)(85001,75.235103)(86001,75.046325)(87001,74.861887)(88001,74.681640)(89001,74.505444)(90001,74.333163)(91001,74.164669)(92001,73.999837)(93001,73.838550)(94001,73.680695)(95001,73.526163)(96001,73.374850)(97001,73.226657)(98001,73.081489)(99001,72.939253)(100001,72.799862)(101001,72.663231)(102001,72.529279)(103001,72.397928)(104001,72.269103)(105001,72.142732)(106001,72.018745)(107001,71.897076)(108001,71.777659)(109001,71.660434)(110001,71.545340)(111001,71.432320)(112001,71.321319)(113001,71.212281)(114001,71.105157)(115001,70.999896)(116001,70.896449)(117001,70.794771)(118001,70.694816)(119001,70.596541)(120001,70.499904)(121001,70.404864)(122001,70.311383)(123001,70.219421)(124001,70.128943)(125001,70.039912)(126001,69.952294)(127001,69.866056)(128001,69.781166)(129001,69.697591)(130001,69.615303)(131001,69.534271)(132001,69.454466)(133001,69.375862)(134001,69.298431)(135001,69.222147)(136001,69.146984)(137001,69.072919)(138001,68.999928)(139001,68.927986)(140001,68.857072)(141001,68.787165)(142001,68.718241)(143001,68.650282)(144001,68.583267)(145001,68.517176)(146001,68.451990)(147001,68.387691)(148001,68.324261)(149001,68.261683)(150001,68.199939)(151001,68.139012)(152001,68.078888)(153001,68.019549)(154001,67.960981)(155001,67.903168)(156001,67.846097)(157001,67.789753)(158001,67.734122)(159001,67.679191)(160001,67.624946)(161001,67.571375)(162001,67.518466)(163001,67.466206)(164001,67.414583)(165001,67.363586)(166001,67.313203)(167001,67.263424)(168001,67.214237)(169001,67.165632)(170001,67.117599)(171001,67.070128)(172001,67.023209)(173001,66.976833)(174001,66.930989)(175001,66.885669)(176001,66.840865)(177001,66.796566)(178001,66.752765)(179001,66.709454)(180001,66.666624)(181001,66.624267)(182001,66.582376)(183001,66.540942)(184001,66.499959)(185001,66.459419)(186001,66.419315)(187001,66.379640)(188001,66.340386)(189001,66.301549)(190001,66.263120)(191001,66.225093)(192001,66.187463)(193001,66.150222)(194001,66.113365)(195001,66.076887)(196001,66.040780)(197001,66.005041)(198001,65.969662)(199001,65.934639)(200001,65.899966)(201001,65.865637)(202001,65.831649)(203001,65.797996)(204001,65.764673)(205001,65.731674)(206001,65.698997)(207001,65.666634)(208001,65.634583)(209001,65.602839)(210001,65.571397)(211001,65.540253)(212001,65.509403)(213001,65.478843)(214001,65.448568)(215001,65.418575)(216001,65.388859)(217001,65.359418)(218001,65.330246)(219001,65.301341)(220001,65.272699)(221001,65.244316)(222001,65.216188)(223001,65.188313)(224001,65.160687)(225001,65.133306)(226001,65.106168)(227001,65.079268)(228001,65.052605)(229001,65.026175)(230001,64.999974)(231001,64.974000)(232001,64.948250)(233001,64.922721)(234001,64.897411)(235001,64.872315)(236001,64.847433)(237001,64.822760)(238001,64.798295)(239001,64.774034)(240001,64.749976)(241001,64.726117)(242001,64.702456)(243001,64.678989)(244001,64.655715)(245001,64.632630)(246001,64.609733)(247001,64.587022)(248001,64.564494)(249001,64.542146)(250001,64.519978)(251001,64.497986)(252001,64.476169)(253001,64.454524)(254001,64.433049)(255001,64.411743)(256001,64.390604)(257001,64.369629)(258001,64.348816)(259001,64.328165)(260001,64.307672)(261001,64.287336)(262001,64.267155)(263001,64.247128)(264001,64.227253)(265001,64.207528)(266001,64.187950)(267001,64.168520)(268001,64.149235)(269001,64.130092)(270001,64.111092)(271001,64.092232)(272001,64.073511)(273001,64.054927)(274001,64.036478)(275001,64.018164)(276001,63.999982)(277001,63.981931)(278001,63.964011)(279001,63.946219)(280001,63.928554)(281001,63.911015)(282001,63.893600)(283001,63.876308)(284001,63.859138)(285001,63.842088)(286001,63.825158)(287001,63.808346)(288001,63.791650)(289001,63.775070)(290001,63.758604)(291001,63.742252)(292001,63.726011)(293001,63.709882)(294001,63.693862)(295001,63.677950)(296001,63.662146)(297001,63.646449)(298001,63.630857)(299001,63.615369)(300001,63.599985)(301001,63.584702)(302001,63.569521)(303001,63.554440)(304001,63.539459)(305001,63.524575)(306001,63.509789)(307001,63.495099)(308001,63.480505)(309001,63.466005)(310001,63.451599)(311001,63.437285)(312001,63.423063)(313001,63.408932)(314001,63.394890)(315001,63.380938)(316001,63.367075)(317001,63.353299)(318001,63.339609)(319001,63.326005)(320001,63.312487)(321001,63.299052)(322001,63.285701)(323001,63.272433)(324001,63.259246)(325001,63.246141)(326001,63.233116)(327001,63.220171)(328001,63.207304)(329001,63.194516)(330001,63.181806)(331001,63.169172)(332001,63.156614)(333001,63.144132)(334001,63.131724)(335001,63.119391)(336001,63.107131)(337001,63.094943)(338001,63.082828)(339001,63.070784)(340001,63.058812)(341001,63.046909)(342001,63.035076)(343001,63.023312)(344001,63.011616)(345001,62.999988)(346001,62.988428)(347001,62.976934)(348001,62.965506)(349001,62.954143)(350001,62.942846)(351001,62.931613)(352001,62.920443)(353001,62.909337)(354001,62.898294)(355001,62.887313)(356001,62.876394)(357001,62.865535)(358001,62.854738)(359001,62.844000)(360001,62.833323)(361001,62.822704)(362001,62.812144)(363001,62.801642)(364001,62.791198)(365001,62.780812)(366001,62.770482)(367001,62.760208)(368001,62.749990)(369001,62.739827)(370001,62.729720)(371001,62.719667)(372001,62.709667)(373001,62.699722)(374001,62.689830)(375001,62.679990)(376001,62.670203)(377001,62.660468)(378001,62.650784)(379001,62.641151)(380001,62.631569)(381001,62.622038)(382001,62.612556)(383001,62.603124)(384001,62.593741)(385001,62.584406)(386001,62.575120)(387001,62.565882)(388001,62.556692)(389001,62.547549)(390001,62.538452)(391001,62.529403)(392001,62.520399)(393001,62.511441)(394001,62.502529)(395001,62.493662)(396001,62.484840)(397001,62.476062)(398001,62.467328)(399001,62.458638)(400001,62.449991)(401001,62.441388)(402001,62.432827)(403001,62.424309)(404001,62.415833)(405001,62.407399)(406001,62.399006)(407001,62.390655)(408001,62.382345)(409001,62.374075)(410001,62.365845)(411001,62.357656)(412001,62.349506)(413001,62.341396)(414001,62.333325)(415001,62.325293)(416001,62.317300)(417001,62.309345)(418001,62.301428)(419001,62.293548)(420001,62.285706)(421001,62.277902)(422001,62.270134)(423001,62.262404)(424001,62.254709)(425001,62.247051)(426001,62.239429)(427001,62.231843)(428001,62.224292)(429001,62.216776)(430001,62.209295)(431001,62.201849)(432001,62.194437)(433001,62.187060)(434001,62.179716)(435001,62.172407)(436001,62.165130)(437001,62.157888)(438001,62.150678)(439001,62.143501)(440001,62.136357)(441001,62.129245)(442001,62.122165)(443001,62.115117)(444001,62.108101)(445001,62.101117)(446001,62.094163)(447001,62.087241)(448001,62.080350)(449001,62.073490)(450001,62.066660)(451001,62.059860)(452001,62.053091)(453001,62.046351)(454001,62.039641)(455001,62.032960)(456001,62.026309)(457001,62.019687)(458001,62.013094)(459001,62.006529)(460001,61.999993)(461001,61.993486)(462001,61.987007)(463001,61.980555)(464001,61.974132)(465001,61.967736)(466001,61.961367)(467001,61.955026)(468001,61.948712)(469001,61.942424)(470001,61.936164)(471001,61.929930)(472001,61.923723)(473001,61.917541)(474001,61.911386)(475001,61.905257)(476001,61.899154)(477001,61.893076)(478001,61.887023)(479001,61.880996)(480001,61.874994)(481001,61.869017)(482001,61.863065)(483001,61.857137)(484001,61.851234)(485001,61.845355)(486001,61.839500)(487001,61.833670)(488001,61.827863)(489001,61.822080)(490001,61.816321)(491001,61.810585)(492001,61.804872)(493001,61.799183)(494001,61.793517)(495001,61.787873)(496001,61.782252)(497001,61.776654)(498001,61.771079)(499001,61.765526)(500000,61.765526)
    };

\addplot+[
    black,very thick,dotted,
    mark=none,
    const plot,
    empty line=jump,
    ]
    coordinates {
    (190,2e6)
    (190,1e2)
    };
  
\legend{Baseline - CPU, Proposed - FiltPIM}
\end{loglogaxis}
\end{tikzpicture}
\caption{Execution time of two platforms running the same filter, the CPU and FiltPIM. Lower latency means better performance.}
\label{fig:results}
\vspace{-12pt}
\end{figure}
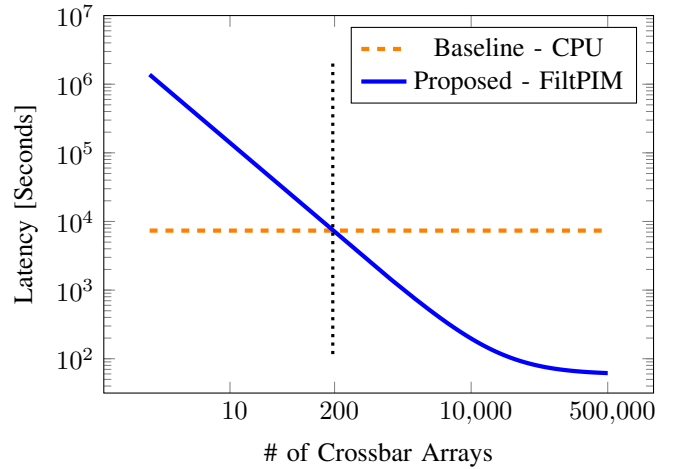

We use the Bitlet model~\cite{bitlet} to evaluate the power consumption of FiltPIM. According to Bitlet, simultaneous operation of 1000 arrays, each using up to 100 rows, consumes 1W. If we set the power budget limit at 100W, only 100K arrays can operate simultaneously. This increases the compute (total) time by 5x (1.75x). Figure~\ref{fig:results} shows how increasing the number of arrays working in parallel in FiltPIM affects its performance. For over 200 working arrays, FiltPIM outperforms the CPU. %When all the 500,000 arrays are working, FiltPIM reaches a speedup of $100x$ over the CPU.

% ---- Conclusion ---- %
\section{Conclusion}
\label{sec:conclusion}

A base count filter improves the performance of DNA read mapping by reducing the work for the computation-intensive sequence alignment stage.
This paper introduces FiltPIM, a novel PIM base-count filter that accelerates filtering by 100x compared to a CPU based base-count filter. This gain is due to reduction in CPU-to-memory data transfer and to the massive parallelism enabled by memristive crossbars. In future work, we aim to build a complete architecture for FiltPIM and measure its benefit to the entire read mapping pipeline.

% ---- Acknowledgment ---- %
\section{Acknowledgment}

This work was supported by the European Research Council through the European Union's Horizon 2020 Research and Innovation Program under Grant 757259.

% Future work??

%\section*{Acknowledgment}
%This work was partially supported by the European Research Council through the European Union's Horizon 2020 Research and Innovation Program under Grant 757259, and partially by the Israel Science Foundation under Grant 1514/17.

% ---- References ---- %
\bibliographystyle{IEEEtran}
\bibliography{refs}

\end{document}